# A high resolution TDC-based board for a fully digital trigger and data acquisition system in the NA62 experiment at CERN

Elena Pedreschi, Bruno Angelucci, Carlo Avanzini, Stefano Galeotti, Gianluca Lamanna, Guido Magazzù, Jacopo Pinzino, Roberto Piandani, Marco Sozzi, Franco Spinella, Stefano Venditti

*Abstract*— **A Time to Digital Converter (TDC) based system, to be used for most sub-detectors in the high-flux rare-decay experiment NA62 at CERN SPS, was built as part of the NA62 fully digital Trigger and Data AcQuisition system (TDAQ), in which the TDC Board (TDCB) and a general-purpose motherboard (TEL62) will play a fundamental role. While TDCBs, housing four High Performance Time to Digital Converters (HPTDC), measure hit times from sub-detectors, the motherboard processes and stores them in a buffer, produces trigger primitives from different detectors and extracts only data related to the lowest trigger level decision, once this is taken on the basis of the trigger primitives themselves.**

**The features of the TDCB board developed by the Pisa NA62 group are extensively discussed and performance data is presented in order to show its compliance with the experiment requirements.**

Manuscript submitted June 26, 2014.

Elena Pedreschi is with Department of Physics, University of Pisa and with the National Institute of Nuclear Research, Pisa, 56127 Italy (telephone: +390502214561, e-mail: elena.pedreschi@pi.infn.it).

Bruno Angelucci is with Department of Physics, University of Pisa and with the National Institute of Nuclear Physics, Pisa, 56127 Italy (e-mail: bruno.angelucci@pi.infn.it).

Carlo Avanzini is with Department of Physics, University of Pisa and with the National Institute of Nuclear Physics, Pisa, 56127 Italy (e-mail: carlo.avanzini@pi.infn.it).

Stefano Galeotti is with the National Institute of Nuclear Physics, Pisa, 56127 Italy (e-mail: stefano.galeotti@pi.infn.it).

Gianluca Lamanna is with the National Institute of Nuclear Physics, Pisa Section, Pisa, 56127 Italy (e-mail: gianluca.lamanna@pi.infn.it).

Guido Magazzù is with the National Institute of Nuclear Physics, Pisa Section, Pisa, 56127 Italy (e-mail: guido.magazzu@pi.infn.it).

Jacopo Pinzino is with Department of Physics, University of Pisa and with the National Institute of Nuclear Physics, Pisa, 56127 Italy (e-mail: jacopo.pinzino@pi.infn.it).

Roberto Piandani is with the National Institute of Nuclear Physics, Pisa Section, Pisa, 56127 Italy (e-mail: Roberto.piandani@pi.infn.it).

Marco Sozzi is with Department of Physics, University of Pisa and with the National Institute of Nuclear Physics, Pisa, 56127 Italy (e-mail: marco.sozzi@pi.infn.it).

Franco Spinella is with the National Institute of Nuclear Physics, Pisa Section, Pisa, 56127 Italy (e-mail: franco.spinella@pi.infn.it).

Stefano Venditti is with CERN PH-ESE-FE Department , 1211 Geneva 23, Geneva, Switzerland (e-mail:stefano.venditti@cern.ch).

## I. INTRODUCTION

THE NA62 experiment at CERN [1] aims at measuring the Branching Ratio (BR) of the ultra-rare Kaon decay $K^+ \rightarrow \pi^+ \nu \bar{\nu}$ [2] as a highly sensitive test of the Standard Model (SM) and a search for New Physics. Since the detection of this process is very challenging due to the smallness of the signal and the presence of a very sizeable background, a very low DAQ inefficiency (below $10^{-8}$) is a key point of the experiment. Several detectors (Fig.1) are placed before, along and after the 65 m long fiducial decay region: some of them are used for $K^+$ and $\pi^+$ tracking, particle identification, photon and muons vetoing, and others for fast timing and trigger.

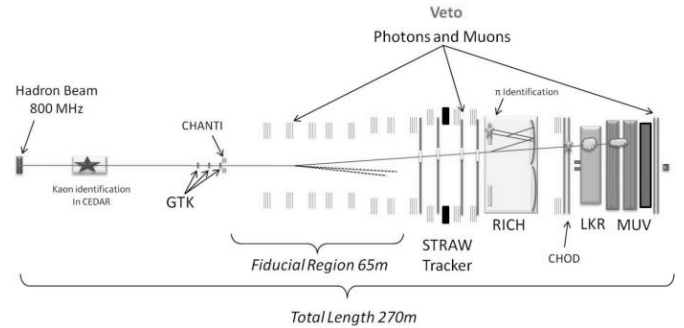

Fig. 1. Schematic view of the NA62 experiment

The design rate for the Level 0 (L0) trigger output is below 1 MHz, with 1ms fixed latency. After L0, data from most detectors is moved to PCs and two further software-based trigger levels are implemented. All electronic boards run on a common, centrally generated, free-running synchronous 40 MHz clock. The L0 trigger is fully digitally implemented, using the same data which is subsequently read out, in order to avoid trigger duplication and data acquisition branches and to allow an accurate offline monitoring. A central trigger processor will asynchronously match L0 trigger primitives generated by a few fast sub-detectors, and dispatch a (synchronous) L0 signal to every board through the same system that distributes the clock to all detectors.

The need for a good time resolution in a high-rate environment, coupled with the desire to have manageable data rates, naturally leads to the use of a TDC-based system where pulse-height information can also be obtained by using a time-over-threshold approach. For these reasons most of the detectors in NA62 adopted TDC-based readout systems.

The devices used for this purposes are a carrier board and a high performance TDC-based daughter-card called respectively TEL62 and TDC Board (TDCB).

## II. THE TEL62 CARRIER BOARD

The data stream in NA62 TDAQ system [3] is completely digital from Front End (FE) to TDAQ and it is based on the TEL62 carrier board [4] which is a major upgrade of the TELL1 board designed by EPFL Lausanne for the LHCb experiment at CERN [5]. The overall architecture of the TEL62 is similar to the TELL1's, but the board has increased internal and external connectivity and is based on much more powerful and modern devices, resulting in eight times the computing power and about twenty times the buffer memory.

The TEL62, used for several sub-detectors in NA62, is a general-purpose data acquisition board and it was chosen as the common underlying support system for NA62 because of its data throughput capabilities, flexibility and the use of a standard commercial data output link. The board is a 9U-format board with connectors for up to four independent daughter-cards (TDCB), each of them served by an FPGA (Altera Stratix III family) and 2GB of dynamic DDR2 RAM (Micron MT16HTF25664HZ-800H1); the four data paths are then connected to a fifth FPGA of the same type which drives a quad GigaBit Ethernet (GBE) daughter card [6] for data transmission; an embedded i386 Credit-Card PC (CC-PC) for slow control (with Ethernet connection) and a TTCrx receiver chip with a Quartz-controlled PLL ASIC (QPLL) are also mounted on the board [7, 8, 11].

The logic of the board is entirely defined by the configuration of the five FPGAs, which can control the other on-board devices (such as TTCrx, RAM and GBE) in a fully custom way.

The main purpose of the TEL62 boards in NA62 is that of pre-processing and monitoring the digital data from sub-detectors, and storing it during the L0 trigger latency period, eventually sending L0-triggered events to PCs. For sub-detectors involved in the formation of the L0 trigger, the same boards would also continuously evaluate trigger primitives on the incoming data, to be sent to the central L0 processor possibly after merging them with those from other boards belonging to the same sub-detector.

## III. TDCB ARCHITECTURE

The TDCB is a high-density mezzanine daughter card for the TEL62 carrier mother-board, designed in Pisa for precision time measurements.

The board design (Fig.2) was driven by the desire to integrate a high number of channels within the same processing board, in order to ease triggering issues. The desire for a compact and common electronics and the relative short distance between the sub-detectors and the place where read-out electronics can be placed, with no space constraints, led to the choice of having digitizers on the readout board rather than on the detector thus leaving only analog front-end electronics on each individual sub-detector (sub-detector specific) in a potentially higher radiation environment and making all digital electronics common and located on the same boards, at the price of having to transmit analog pulses on the 5 m LVDS cables between the two.

The requirements of a good time resolution and high channel integration led to the choice of the CERN High Performance Time to Digital Converter (HPTDC or TDC) as time digitizers [9].

The HPTDC can work in an un-triggered mode, in which all available data is delivered at every readout request, or in a trigger-matching mode, in which a readout request follows a trigger pulse, and the TDC only delivers the data which matches in time the trigger occurrence, within some programmable time windows. Trigger-matching mode was implemented to allow HPTDCs to work as front-end buffers, storing data while a trigger signal was generated; however, in a modern experiment such as NA62, the latency of the lowest trigger level is much longer than the typical time it takes to fill on-chip buffer. The TDCBs therefore normally use TDCs in trigger-matching mode just as a way of obtaining properly time-framed data, but triggers are actually sent to HPTDCs in a continuous periodic stream, with no relation whatsoever to the trigger of the experiment (data storage during trigger latency being actually provided on the carrier board with much larger buffers). The time-matching parameters in the TDCs have to be properly set in order to allow readout of all hits which occurred since the previous trigger (time-matching window set equal to trigger period): in this way the TDCs are periodically triggered and readout, receiving all hits corresponding to a time frame of length equal to the triggering period, in a continuous sequence. Since the range of time measurement within the TDC chips is limited (to 51.2 μs at most), and therefore a roll-over of the TDC time word frequently occurs, only by exploiting this working mode one can be guaranteed (by the TDCs themselves) that all the data are being read.

Each HPTDC provides 32 TDC channels when operated in fully digital mode at 98 ps LSB resolution, with some internal buffering for multi-hit capability and a trigger-matching logic allowing the extraction of hits in selected time windows. Four such chips, for a total of 128 channels, are mounted on each TDCB, resulting in a grand total of 512 TDC channels per fully-equipped TEL62 carrier-board; as an example, the entire RICH detector can be handled by 4 TEL62 boards only, while most small sub-detectors only require a single TEL62 board.

With a fast front-end electronics providing adequately time-stretched LVDS discriminated pulses, the measurement of both the leading and trailing edge times allows obtaining analog pulse-height information by the time-over-threshold method: HPTDCs can indeed digitize the time of occurrence of both signal edges, provided they are separated by a minimum time (7 ns); the resulting time measurements (made with respect to clock edges) can be combined into a single word to reduce the required data bandwidth.

The board houses four 68-pin VHDCI connectors for input signals, each of them delivering 32 LVDS signals to one TDC,

with two spare pairs being used to provide additional grounding (one pair) and to allow user-defined back communication from the TDCB to the front-end electronics (one pair).

This latter feature can be used to trigger the injection of calibration pulses in the sub-detector or calibration patterns in the front-end (as two single-ended lines allowing bidirectional communication, or as a LVDS pair towards the front-end). This choice allows in principle the use of high-performance cables, if required by the intrinsic resolution of a sub-detector, as well as cheaper solutions.

The TDCB houses a dedicated FPGA (Altera Cyclone III family), named TDC Controller (TDCC-FPGA) which can handle the configuration of the four HPTDCs via JTAG, read the data they collect, and possibly pre-process it. A 2MBs external static RAM block is also available and will be used for online data monitoring purposes, and low-level checks on data quality.

The TDCC-FPGA can be configured from an on-board flash memory (Altera EPCS64SI16N), which can be loaded using an external programmer via an on-board connector or through JTAG, either using a JTAG port on the TEL62 board, or through its CC-PC card. Communication between each TEL62 FPGA and the corresponding TDCC-FPGA on the daughter-board proceeds through a 200-pin connector with 4 independent 32-bit single-ended LVTTL parallel data buses (one for each TDC) and a few dedicated lines for synchronous commands and resets. The TDCC-FPGA on the TDC daughter-card can also be accessed from the TEL62 CC-PC card via a dedicated I2C connection for slow operations. The individual TDCs are configured via JTAG, with the TDCC-FPGA acting as the JTAG master: the configuration bits are sent to the TDCC-FPGA from the TEL62 CC-PC card via I2C, and are then uploaded to the TDCs. A second working mode allows inserting both the TDCC-FPGA and the four HPTDCs into a global JTAG chain which also includes all the TEL62 devices, and which can be driven by the TEL62 CC-PC card.

The contribution of the digitizing system to the time resolution ultimately depends on the random jitter of the reference clock against which the measurement is performed. The 40 MHz clock is optically distributed by the TTC system [10] to each TEL62 board, where it is cleaned by the QPLL [11] to reduce the jitter below 50 ps. This clock signal drives the internal logic and is distributed to each TDCB, where it can be configured to go through two more jitter-cleaning stages: these are the internal PLL of the on-board TDCC-FPGA and a second on-board QPLL [11].

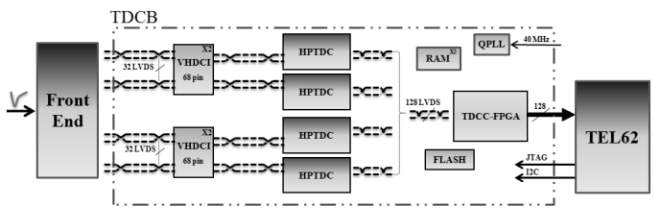

Fig. 2. TDCB Architecture's block diagram

## IV. TDCB FIRMWARE

A firmware (Fig.3) for the TDC board's TDCC-FPGA has been developed by the NA62 Pisa group [12]. In the configuration chosen by NA62, the channel within the TDCs will produce two 32 bit-long words for each LVDS signal in each channel, one word for the leading and one for the trailing edge. 19 bits out of 32 are dedicated to time measurement, while the other bits identify the channel (5 bits), the TDC inside the board (4 bits) and the type of data sent (4 bits), i.e. leading or trailing edge, error words and TDC frame timestamps. Through a 32-bit parallel block writing protocol the TDCC-FPGA receives the data sent by the HPTDCs and performs several operations on them before they are addressed to the TEL62 carrier board:

### A. Appending a time stamp and a word counter to the data stream from the TDCs

The TDC data format is such that the time measurement rolls over every 51.2 μs; therefore a timestamp should be added to data in such a way that the TDC time measurement is unambiguously associated to an "absolute" time which can cover a whole accelerator spill (up to a few seconds). This is achieved by periodically triggering the TDCs with a period shorter than the TDC's rollover (presently a trigger is sent by the TDCC-FPGA every 6.4 μs) and adding a frame timestamp at the beginning of the data stream associated with each trigger. A word counter, indicating the number of words received by the TDC and the frame timestamp, is also appended at the end of the frame data stream. The rollover of the frame timestamp (whose LSB corresponds to 400 ns) is long enough to cover foreseen spill durations.

### B. Configuring TDCs and communicating with the CC-PC Card

As mentioned before, configuration data is sent to the TDCB through I2C. An I2C slave controller has been implemented in the firmware, through which internal registers are both written and read and a JTAG master controller is therefore implemented in the TDCC-FPGA. Data to/from the TDCs are transmitted/received through the JTAG protocol, to which the I2C controller interfaces.

### C. Emulator

Two different emulators are contained in the TDCB firmware: one suited to send some simple repeating pattern on selected TDC channels (single-channel TDC emulator), and one which allows to fully specify some amount of data words which are loaded from a file into memory and repeatedly sent (data emulator).

In particular the data emulator generates data packets containing a variable number of words per frame as required by the user, sending one data packet for each frame to which the header (timestamp) and trailer (word count) are automatically added; these data words are inserted in the output FIFO directly.

The data words are read from a 32-bit wide and 1K deep pattern memory, and modified so that the upper bits of the TDC time field do match the current frame timestamp (otherwise the words might be rejected in later stages of the processing): this means that the data words will not repeat exactly and will not be exactly as loaded in the pattern memory. The number of words to be put in each frame are read from each line of a 9-bit wide and 1K deep count memory: a 0 value will generate an empty frame (only header and trailer), while other numbers will generate a frame with that number of data words (as read from the pattern memory); note that there is a limit on the number of words per frame, which should not exceed 0xFA (250 words), and a frame must be completed before a new trigger comes (with real TDC data this is not a strict requirement and it can happen that an outstanding frame will take longer than the inter-trigger period to be transferred, but this is not currently implemented in the data emulator). When the last address of the counter memory is reached the first one is read again, repeating the same sequence. The upper bit in the count memory (e.g. a count value 0x100) forces a rewind so that the next event will be as the first one.

In order to fill the pattern and count memories some suitable I2C registers are used to implement an indirect memory addressing.

### D. Enabling RAM reading/writing

During the acquisition, the data stream can be optionally split and sent to the TDCB's on-board RAM.

The RAM space is equally subdivided among the four TDCs. In the final implementation, it will be possible to choose between filling the RAM space with the first data of the spill or collecting only the most relevant data, for example that corresponding to events where one or more TDCs sent an error word. This data (which will not necessarily be collected through the main data stream) can be then analyzed off-line to better study some fine details of the data for debugging or monitoring purposes.

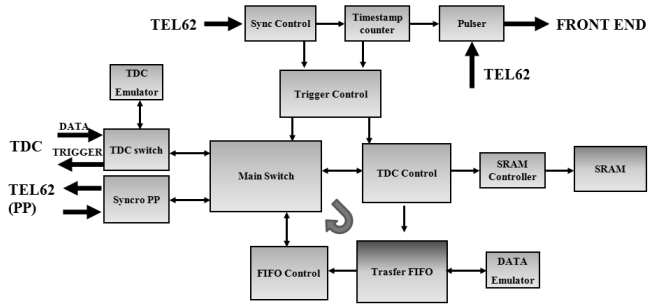

Fig. 3. TDCB Firmware's block diagram

## V. TDCB TESTS AND RESULTS

Once the TDCB design was finalized various tests were performed in order to verify that the system complies with the experiment's requirements. Some of such tests were performed in laboratory using a test setup which may reproduce conditions similar to the experiment environment, others were performed at CERN during a test beam.

### A. Time Resolution and Cross-Talk

A measurement of the intrinsic contribution of the board to the time resolution was evaluated in the laboratory by pulsing TDC channels with signals of fixed duration generated by an FPGA-based test board working on the same clock used by the TDCB, and measuring the time difference between the trailing and leading times of a LVDS pulse with nominal 25 ns duration.

When pulsing 32 TDC channels at a time we observed that the RMS of the measured pulse width on a single channel was 0.061 ns (Fig. 4), with an average of 25.18 ns, for $10^5$ pulses. The values are comparable when only 1 every 4 TDC channels (8 in total) is pulsed simultaneously with the same signals of the previous case (Fig. 5).

No spurious hits were detected on channels which were not pulsed with a statistic of $10^7$ events, but the effect of possible cross-talk was checked by comparing a channel time resolution into 2 different cases: al the 32 TDC channels pulsed with signals of constant width (25 ns) or pulsing only 8 TDC channels (1 every 4); since no differences are observed in the time resolution (see Fig. 4 and 5) we conclude that no significant effect of cross talk is present.

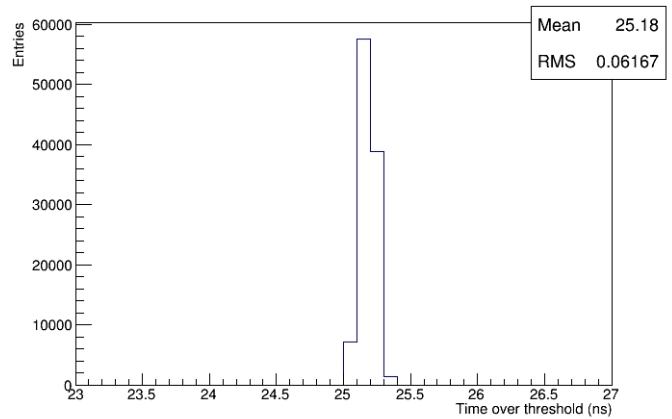

Fig. 4. TDCB Time Resolution: 32 channels pulsed with 25 ns signal

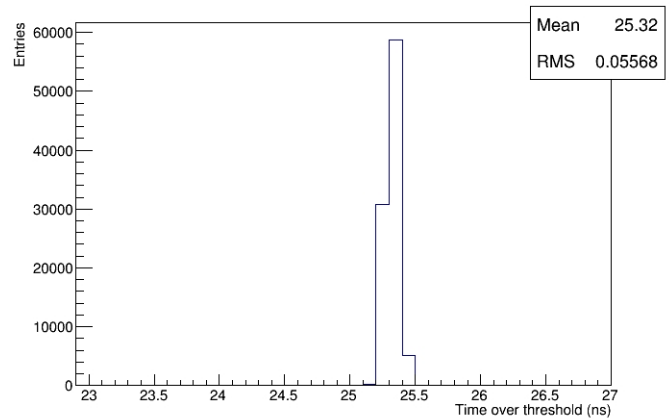

Fig. 5. TDCB Time Resolution: 8 channels pulsed with 25 ns signal

### B. Channel Efficiency

The data transfer efficiency (epsilon) is defined as the ratio between the number of hits delivered by the system and the number of input pulses, and was measured in the laboratory both on a single channel (1 of every group of 8 for a total of 4 TDC channels) and on two neighboring ones (channels 0 and 1 of every group of 8 for a total of 8 TDC channels) (Fig. 6)

No data losses were observed for input rates below 17.5 MHz in the first case (single channel).

In the case of two adjacent channels being pulsed, no data losses were observed for input rates below 8.5 MHz per channel; above such value data losses start to appear at slightly different levels depending on the pulsed channel.

Hit losses are expected at high rates due to the limited amount of buffering present in the HPTDC.

In the HPTDC the 32 channels are divided in four independent groups of 8 channels each, and hits are buffered at different stages: per channel, per group of 8 channels, per full chip; apart from the first buffering stage, the other ones are monitored for overflow conditions.

The first channels in each group of 8 are serviced faster than the last ones due to the simple fixed-priority arbitration scheme adopted. At high rates this gives an advantage to the low-numbered high-priority channels which can use their channel de-randomizers more efficiently, while in such conditions lower-priority channels appear like they have smaller de-randomizing capability and therefore slightly higher data losses [9].

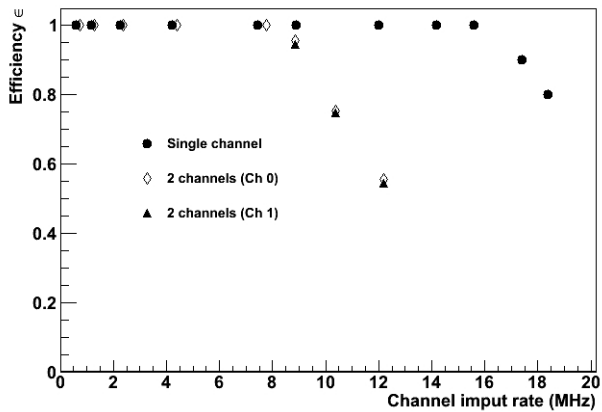

Fig. 6. TDCB Channel Efficiency for one and for two channels

### C. Test Beam Results

In November and December 2012 a test run with beam was performed at the CERN SPS. The response of the detector, the time and space correlations between the detectors, the rates and efficiencies of the detectors and the TDAQ system were the main goals of this test.

As in the experiment for the test beam the secondary Kaon beam was produced by protons from the SPS at 400 GeV/$c$ impinging on a beryllium target; the beam optics system selected the charged Kaons with a momentum of ($75 \pm 1\%$) GeV/$c$. The choice of a relatively high-energy beam helps in background rejection and sets the longitudinal scale of the experiment. The detectors are positioned along 170 m, starting about 100m after the beryllium target; the fiducial region for the selection of useful decays was 65m long. The total nominal beam rate for the experiment is 750 MHz, of which charged Kaons are about 6%, resulting in an integrated event rate over downstream detectors of about 10 MHz.

The test was carried out with a much reduced intensity (1/50 of nominal).

The event time difference distribution between the fast and high-resolution detectors CHOD and CEDAR (Fig. 7) had a fitted standard deviation of 410 ps, a value which is fully compatible with the intrinsic detector resolutions, thus confirming a negligible contribution from the electronics.

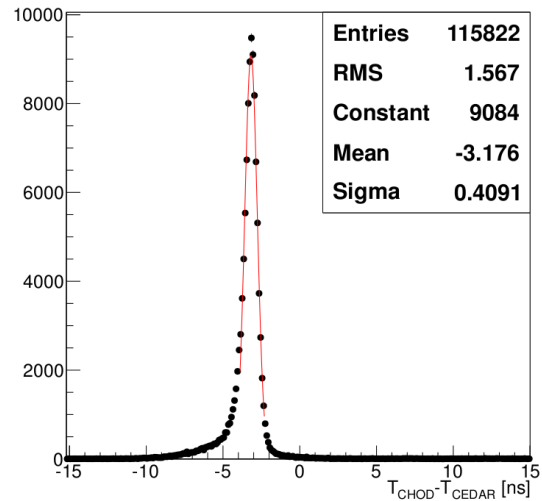

Fig. 7. Event time difference between CHOD and CEDAR detectors with Gaussian fit.

## VI. CONCLUSIONS

Data collected in the laboratory and during a test beam confirmed that the system complies with the experiment requirements and showed that the time resolution is adequate and compatible with the expected time resolution of each sub-detector

The TDCB board developed by the NA62 Pisa group provides a compact, reliable and flexible solution for the digitization of signals from the majority of the detectors in NA62, a new high-rate particle physics experiment. The HPTDCs used in a continuous triggering mode were proven to be a reliable solution to collect data used in the L0 trigger and to cope with the very low inefficiency required by the NA62 experiment in order to reach its physics goal.